\documentclass{article}

\usepackage{arxiv}
\usepackage[utf8]{inputenc} 
\usepackage[T1]{fontenc}    
\usepackage{hyperref}       
\usepackage{url}            
\usepackage{booktabs}       
\usepackage{amsfonts}       
\usepackage{nicefrac}       
\usepackage{microtype}      
\usepackage{lipsum}
\usepackage{graphicx}
\usepackage{subcaption}
\usepackage{amsmath,amsfonts,amssymb}
\graphicspath{{Figures/}}

\newcommand{\RM}[1]{#1}

\title{Non-autoregressive time-series methods for stable parametric reduced-order models}

\author{
  Romit Maulik \\
  Argonne Leadership Computing Facility\\
  Argonne National Laboratory\\
  Lemont IL 60439, USA\\
  \texttt{rmaulik@anl.gov} \\
  \AND
  Bethany Lusch \\
  Argonne Leadership Computing Facility\\
  Argonne National Laboratory\\
  Lemont IL 60439, USA\\
  \texttt{blusch@anl.gov} \\
  \AND
  Prasanna Balaprakash \\
  Mathematics and Computer Science Division \& \\
  Argonne Leadership Computing Facility \\
  Argonne National Laboratory\\
  Lemont IL 60439, USA\\
  \texttt{pbalapra@anl.gov} \\
}

\begin{document}
\maketitle

\begin{abstract}
Advection-dominated dynamical systems, characterized by partial differential equations, are found in applications ranging from weather forecasting to engineering design where accuracy and robustness are crucial. There has been significant interest in the use of techniques borrowed from machine learning to reduce the computational expense and/or improve the accuracy of predictions for these systems. These rely on the identification of a basis that reduces the dimensionality of the problem and the subsequent use of time series and sequential learning methods to forecast the evolution of the reduced state. Often, however, machine-learned predictions after reduced-basis projection are plagued by issues of stability stemming from incomplete capture of multiscale processes as well as due to error growth for long forecast durations. To address these issues, we have developed a \emph{non-autoregressive} time series approach for predicting linear reduced-basis time histories of forward models. In particular, we demonstrate that non-autoregressive counterparts of sequential learning methods such as long short-term memory (LSTM) considerably improve the stability of machine-learned reduced-order models. We evaluate our approach on the inviscid shallow water equations and show that a non-autoregressive variant of the standard LSTM approach that is bidirectional in the PCA components obtains the best accuracy for recreating the nonlinear dynamics of partial observations. Moreover---and critical for many applications of these surrogates---inference times are reduced by three orders of magnitude using our approach, compared with both the equation-based Galerkin projection method and the standard LSTM approach. 
\end{abstract}

\section{Introduction}
\label{intro}

Recently, researchers have shown sustained interest in using machine learning methods for bypassing traditional numerical methods \cite{long2017pde,raissi2018deep,raissi2019physics,sirignano2018dgm}. This is due to the promise such methods hold in multiple applications ranging from engineering design to climate modeling, where forward model solves rely on nonlinear partial differential equations (PDEs). Frequently,  these systems exhibit multiscale and advective behavior, which leads to very fine spatiotemporal discretization requirements. Consequently, PDE-based solutions of these systems become computationally expensive, causing a significant bottleneck in design and forecast tasks \cite{verstappen1997direct}. Data-driven reduced-order methods (ROMs) are promising since they allow for rapid predictions of nonlinear dynamics unencumbered by the limitations of numerical discretizations \cite{raissi2019physics,wang2019towards}. In almost all ROM applications, forecasts must be conditioned on time and several control parameters such as the initial conditions or the physical properties of the governing laws. Moreover, most systems need to be integrated in time for a large duration; and stable predictions throughout the lifetime of the dynamical system are essential. We address issues related to the stability of conditional surrogates by introducing physics-informed non-autoregressive methods for time series prediction. To that end, we propose a novel long short-term memory (LSTM) network method that performs bidirectional \cite{graves2005framewise} gating in the dimension of the principal component analysis (PCA) coefficients while being globally connected in time. This method is compared with standard techniques such as the traditional  LSTM \cite{hochreiter1997long}, a non-autoregressive version of a temporal convolutional network \cite{oord2016wavenet}, and a non-autoregressive multilayered perceptron. Our testing results show lower testing and reconstruction errors from the proposed method as well as significant improvement in model stability compared with that of the traditional LSTM. Assessments are also made against the Galerkin projection (GP) \cite{rowley2004model}, and our proposed approach is seen to provide more accurate results with shorter inference times. To summarize, the contributions of this article are as follows:
\begin{itemize}
    \item A novel non-autoregressive LSTM-based method is proposed that performs bidirectional gating in the PCA dimension while remaining fully connected in time for predicting the spatiotemporal dynamics of the shallow water equations.
    \item We demonstrate that the proposed method can provide more accurate results for learning the trajectory of the nonlinear dynamical systems in addition to providing much lower inference times and exhibiting greater stability.
    \item We advance the state of the art for forecasting nonlinear dynamical systems from the point of view of stability and the ability to handle incomplete data.
\end{itemize}


\section{Related work}
\label{Related}

Neural networks have been used for ROMs for decades. One of the earliest examples \cite{hsieh1998applying} used a simple fully connected network for forecasting meteorological information.  More recently, researchers  have incorporated a single-layered feed-forward neural network into a nonlinear dynamical system and built a surrogate model for a high-dimensional aerodynamics problem \cite{mannarino2014nonlinear};  radial basis function networks have been used to make forecasts for a nonlinear unsteady aerodynamics task \cite{zhang2016nonlinear,kou2017layered}; and  a simple fully connected network has been used for learning the dynamics of an advection-dominated system \cite{san2018neural,hesthaven2018non}. Data generated from PDE simulations can often be interpreted as images on a square grid, so convolutional neural networks have also been applied \cite{thuerey2019deep,kim2019deep}. We call this category of models \emph{nonintrusive} since they can be obtained solely from the data.

Although  other ways can be used to reduce the dimensionality of dynamical systems data, using the PCA projection means that the latent space can be interpreted as evolving coefficients in terms of physically relevant spectral content. Since 2018, there has been major growth in the use of LSTMs after projecting dynamical systems into latent PCA space \cite{wang2018model,wan2018data,mohan2018deep,rahman2019nonintrusive,deng2019time,maulik2020time}. Since errors from the neural network model accumulate, the autoregressive approach may be unstable for long-term predictions. Most studies do not report on the robustness and stability of their trained networks for long prediction horizons. 

In addition, such studies assume that all dependent variables are observable. This assumption limits the application of these methods to realistic forecasting scenarios where not all of the physical processes are observable. In practice,  training recurrent networks for adhering to physical manifolds is nontrivial \cite{pearlmutter1989learning}, and it is made doubly difficult by having incomplete access to all the relevant information. Recent articles have tried to address these limitations \cite{greydanus2019hamiltonian} by constructing physics-aware networks that learn conservation laws, but their extension to complicated systems with incomplete observations remains unclear.  

The method proposed in this article represents an improvement in the state of the art for nonintrusive ROMs based on recurrent neural networks. We find that a non-autoregressive approach improves long-term stability and inference time. It also allows a novel use of an LSTM where gating is performed in the PCA dimension instead of the time dimension. The error is further decreased by using a bidirectional LSTM. We perform a thorough analysis of the stability and robustness of the proposed framework, and we find improved performance in comparison with that of traditional methods. We also assess the performance of the framework for incomplete observations. GP, an \emph{intrusive} and equation-based method \cite{rowley2004model} is used as a baseline for the purpose of an additional comparison. We note that GP requires the solution of a partial differential equation in latent space and complete observations of the system dynamics. We demonstrate that our proposed method, which operates on incomplete observations, outperforms GP as well.

\section{Methods}
\label{Methods}

\begin{table}[ht]
\small
\centering
\caption{Methods starting with ``A-'' are those where outputs are fed back into the network for recursive prediction, ``NA-'' methods directly forecast all dynamics at once, and ``T/P'' indicates which dimension is being gated (time/PCA, respectively). The last column is the number of outputs from each network architecture during training.}
\begin{tabular}{|c|c|c|c|}
\hline
Method & Gating& Gating   & Prediction \\ 
 & Space & Type  & Steps\\ \hline
\multicolumn{4}{|c|}{Autoregressive} \\\hline
A-LSTM-T               & Time         & Standard & 1        \\ \hline
A-LSTM-T-R             & Time         & Standard & 2        \\ \hline
\multicolumn{4}{|c|}{Non-autoregressive} \\\hline
NA-LSTM-T               & Time         & Standard & $\mathbf{N-k}$      \\ \hline
NA-TCN                   & N/A          & N/A      & $\mathbf{N-k}$      \\ \hline
NA-LSTM-P               & PCA          & Standard & $\mathbf{N-k}$      \\ \hline
NA-BLSTM-P             & PCA          & Bidirectional & $\mathbf{N-k}$ \\ \hline
NA-MLP                & N/A          & N/A           & $\mathbf{N-k}$ \\ \hline
\multicolumn{4}{|c|}{Analytical} \\\hline
GP            & N/A          & N/A      & 1        \\ \hline
\end{tabular}
\label{Table1}
\end{table}

The parameterized forecasting of a high-dimensional advection-dominated problem can be formulated as a supervised sequential learning problem. A common approach for solving this problem comprises four steps:
\begin{enumerate}
    \item Collect time series data $\mathbf{u_1, u_2, \dots, u_N} \in \mathbb{R}^m$, evenly spaced in time. For example, the data could be from a spatiotemporal system governed by a PDE. However, we do not assume that we have ``complete information'' in the sense of measuring all  the dependent variables of the original system.
    \item Reduce the dimensionality of the problem by projecting data into the latent space defined by the first $\mathbf{r}$ PCA components of the data, producing $\mathbf{z_1, z_2, \dots, z_N} \in \mathbb{R}^\mathbf{r}$.
    \item Train a time series model in this latent space. The inputs are a historical sequence of inputs $\mathbf{z_{n-k}},\mathbf{z_{n-k+1}},\hdots, \mathbf{z_{n}}$ and control variables $w$, such as the initial conditions. A trained model is tasked with predicting the sequence of outputs $\mathbf{z_{n+1}}, \mathbf{z_{n+2}}, \hdots, \mathbf{z_{n+T}}$, where $\mathbf{T}$ is the total number of forecast steps.
    \item Project the predictions $\mathbf{z_t}$ at any future timestep $\mathbf{t}$  back to $\mathbb{R}^m$ using the saved PCA bases to assess reconstruction fidelity.
\end{enumerate}

\subsection{Galerkin Projection}

For comparison of all  our machine-learned predictive strategies, we utilize the Galerkin projection  \cite{juang1985eigensystem,hall2000proper} methodology, which has been used extensively  for advection-dominated systems including the shallow water equations \cite{cstefuanescu2013pod,hijazi2019data}. Briefly, GP involves the projection of the governing partial differential equations onto the truncated PCA space. The orthonormality of the PCA bases leads to a significant reduction in the number of coupled ordinary differential equations (for instance, the retention of $\mathbf{r}$ basis vectors implies that $\mathbf{r}$ coupled systems must be solved). However,  some severe limitations are associated with this approach that hamper its utility as a surrogate model for dynamical systems. First, the utilization of GP necessitates complete observation of all dependent variables in the system. Second, the projection of the governing equations to the PCA space leads to a drop in accuracy since higher-order interactions between the truncated PCA bases are lost. Third, when the number of retained components grows, the computational cost of GP becomes prohibitive \cite{chaturantabut2010nonlinear}. These factors have contributed to growing interest in the use of machine learning methods for time series data for bypassing equation-based surrogates. GP results  serve as a benchmark comparison for our proposed method against a state-of-the-art analytical technique and a full formulation of the same for this test case may be found in our previous work\cite{maulik2020reduced}.

\subsection{Time-series learning methods}

 Here, we introduce some common methods for the prediction of time series data, such as the LSTM network, and we describe our proposed non-autoregressive adaptation to these methods to achieve our goal of stable and accurate predictions on test datasets.  Table \ref{Table1} shows the nomenclature adopted in this study for the various time-series learning methods.

\subsubsection{Autoregressive methods}

Our first method is the traditional LSTM network \cite{hochreiter1997long}. In this method, gating is applied in the time dimension, and the framework seeks to predict the time-varying coefficients of the PCA bases. In addition, the outputs of the network are fed back into the window required for predicting the next step. This \emph{sliding-window} prediction strategy is denoted \emph{autoregressive}. A vast majority of nonintrusive surrogate modeling strategies employ this methodology for one-step-ahead prediction of dynamics. The choice is motivated primarily by the methodological similarities with most ordinary differential integration techniques \cite{dormand1980family}, which integrate dynamics one step at a time because of issues of \emph{numerical} stability. We denote this method A-LSTM-T (short for autoregressive LSTM with gating in time and prediction in PCA space). \RM{The evolution equations for a standard LSTM are given as follows:
\begin{align}
\begin{split}
\text{input gate: }& \boldsymbol{G}_{i}=\boldsymbol{\varphi}_{S} \circ \mathcal{F}_{i}^{N_{c}}(\mathbf{z_n}), \\
\text{forget gate: }& \boldsymbol{G}_{f}=\boldsymbol{\varphi}_{S} \circ \mathcal{F}_{f}^{N_{c}}(\mathbf{z_n}), \\
\text{output gate: }& \boldsymbol{G}_{o}=\boldsymbol{\varphi}_{S} \circ \mathcal{F}_{o}^{N_{c}}(\mathbf{z_n}), \\
\text{internal state: }& \boldsymbol{s}_{n}=\boldsymbol{G}_{f} \odot \boldsymbol{s}_{n-1}+\boldsymbol{G}_{i} \odot\left(\boldsymbol{\varphi}_{T} \circ \mathcal{F}_{is}^{N_{c}}(\mathbf{z})\right), \\
\text{output: }& \mathbf{h}_{n+1} = \boldsymbol{G}_{o} \circ \boldsymbol{\varphi}_{T}\left(\boldsymbol{s}_{n}\right),
\end{split}
\end{align}
where $\mathbf{z_n}$ is an input at a current time step and $\mathbf{a} \circ \mathbf{b}$ refers to a Hadamard product of two vectors. The above set of operations are \emph{unrolled} in the temporal dimension to allow for the effect of $\mathbf{z_{n-k}}, \mathbf{z_{n-k+1}}, \hdots, \mathbf{z_{n-1}}$ on $\mathbf{z_{n}}$ for making a prediction for $\mathbf{z_{n+1}}$. Note that, $\boldsymbol{\varphi}_{S}$ and $\boldsymbol{\varphi}_{L}$ refer to tangent sigmoid and tangent hyperbolic activation functions, respectively, and $N_c$ is the number of hidden layer units in the LSTM network. Also, $\mathcal{F}^{n}$ refers to a linear operation given by a matrix multiplication and subsequent bias addition, i.e.,
\begin{align}
\mathcal{F}^{n}(\boldsymbol{x})=\boldsymbol{W} \boldsymbol{x}+\boldsymbol{B},
\end{align}
where $\boldsymbol{W} \in \mathbb{R}^{n \times m}$ and $\boldsymbol{B} \in \mathbb{R}^{n}$ for $\mathbf{x} \in \mathbb{R}^m$. Conventionally, the output of the standard LSTM $\mathbf{h_{n+1}}$ may fed into another set of LSTM operations if multiple cells are stacked. If there is only one cell or these operations represent those for the final one, the output is acted upon by another operation akin to a linear operation followed by tangent sigmoid activation to obtain $\mathbf{z_{n+1}}$, i.e., 
\begin{align}
    \mathbf{z_{n+1}} = \varphi_T (\mathcal{F}_{op}^\mathbf{r} \mathbf{h_{n+1}}).
\end{align}
}
A key disadvantage of the autoregressive time series models is that they suffer from problems related to error propagation during recursive predictions. This often results from a lack of an \emph{analytical} notion of stability \cite{maulik2020time}. Therefore, we consider a modified version of the A-LSTM-T method where the training involves a metamodeling strategy allowing for output feedback in the training process. The computational graph is set up in such a way that any error propagation due to prediction feedback is penalized. \RM{This may be represented as 
\begin{align}
    \begin{gathered}
    \mathbf{z_{n+1}} = \mathcal{M}_{A} (\mathbf{z_{n-k}}, \mathbf{z_{n-k+1}}, \hdots, \mathbf{z_{n}}) \\
    \mathbf{z_{n+2}} = \mathcal{M}_{A} (\mathbf{z_{n-k+1}}, \mathbf{z_{n-k+2}}, \hdots, \mathbf{z_{n+1}})
    \end{gathered}
\end{align}
where $\mathcal{M}_{A}$ is an aggregation of the LSTM operations and the computation of $\mathbf{z_{n+2}}$ depends on the \emph{prediction} of $\mathbf{z_{n+1}}$ by the current state of the network.} We note that this technique has previously been utilized for training latent space representations for nonlinear dynamical systems \cite{lusch2018deep}. We assess whether it can enhance the stability of the standard A-LSTM-T. We denote this method as A-LSTM-T-R. Both A-LSTM-T and A-LSTM-T-R require a window of inputs to make a one-step prediction. In addition, parameter information (i.e., $w$) is concatenated to this window-augmented state vector. To summarize, our training dataset for the autoregressive methods has multiple examples of inputs of a window of state vectors. The output from these methods is a forecast of the state of the dynamical system at the next timestep.

\subsubsection{Non-autoregressive methods}

\begin{figure}[h!]
    \centering
    \includegraphics[width=0.9\textwidth]{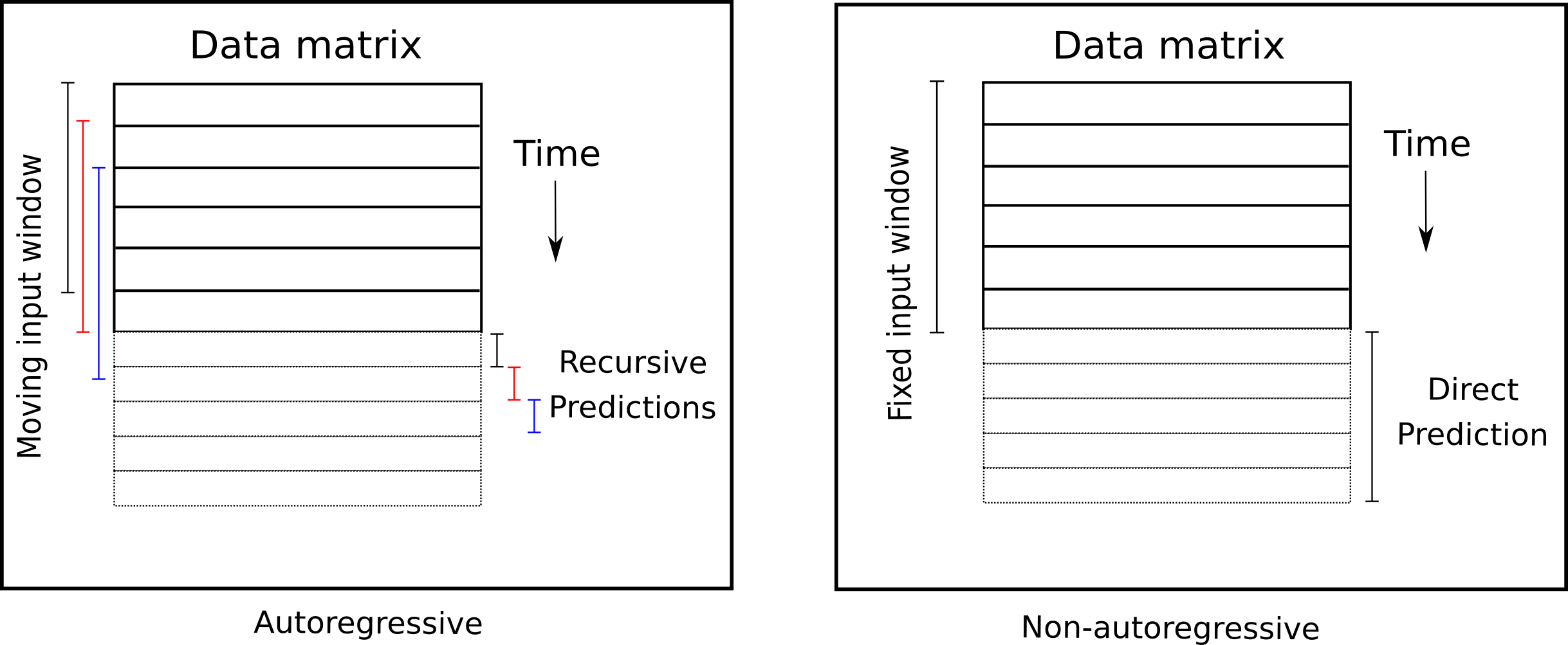}
    \caption{Schematic outlining the difference between autoregressive and non-autoregressive methods for forecasting. The solid rectangles in each data matrix indicate the state of the dynamical system and their dotted versions indicate states that need to be predicted. Non-autoregressive methods have been proposed to deal with issues of noise accumulation in regular time-series forecasting techniques.}
    \label{Gating_Schematic}
\end{figure}

Here, we present non-autoregressive methods that help improve the stability and accuracy of surrogate models. For this reason, the methods  are prepended with ``NA.'' \RM{The equations of a general non-autoregressive method are given by a \emph{direct prediction} as follows,
\begin{align}
    \mathbf{z_{n+1},z_{n+2},\hdots,z_{n+T}} = \mathcal{M}_{NA} (\mathbf{z_{n-k},z_{n-k+1},\hdots,z_{n}})
\end{align}
where $\mathcal{M}_{NA}$ is a non-autoregressive map. A schematic outlining the difference between autoregressive and non-autoregressive methods is provided in Figure \ref{Gating_Schematic}.}

Our first approach is to introduce the NA counterpart of the A-LSTM-T method. Here, a standard LSTM is configured to return a sequence of all forecasts, given a burn-in sequence of $\mathbf{k}$ inputs and parameter information. A direct prediction of the dynamics precludes the necessity for any autoregressive feedback. \RM{We denote this method NA-LSTM-T and the sole difference from the auto-regressive methods of the previous section are due to the final operation on the LSTM output, i.e.,
\begin{align}
    \mathbf{z_{NA}} = \varphi_T (\mathcal{F}_{op}^{\mathbf{r} \times \mathbf{N-k}} \mathbf{h_{n+1}}),
\end{align}
where $\mathbf{z_{NA}}$ is a vector that stacks the states at different time steps into one target. Effectively, we predict all future states at once by managing the output dimension of the standard LSTM.}

More importantly, the non-autoregressive formulation allows one to explore alternative strategies in terms of the interpretation of the dataset. For instance, we may interpret the dataset to be \emph{sequential in PCA space} rather than time. A framework leveraging this interpretation can be devised simply by switching the gating dimension of the dataset. This aligns with the sequential nature of the PCA coefficients in terms of the proportion of variance capture of the dataset. The first method that leverages this is thus called NA-LSTM-P. \RM{The equations of the NA-LSTM-P method are given as follows. First let us consider a matrix $\mathbf{Z} \in \mathbb{R}^{\mathbf{k} \times \mathbf{r}}$ where each row corresponds to the $\mathbf{r}$-dimensional reduced state at each time step $k$ of the burn-in window $\mathbf{k}$. We may transpose this matrix to obtain $\mathbf{Z}^{\prime} \in \mathbb{R}^{\mathbf{r} \times \mathbf{k}}$ where each row now corresponds to the PCA coefficient $r$ (of a total of $\mathbf{r}$ coefficients) of our reduced state. Our NA-LSTM-P method would then perform the following operations,
\begin{align}
\begin{split}
\text{input gate: }& \boldsymbol{G}_{i}=\boldsymbol{\varphi}_{S} \circ \mathcal{F}_{i}^{N_{c}}(\mathbf{z^\prime}_r), \\
\text{forget gate: }& \boldsymbol{G}_{f}=\boldsymbol{\varphi}_{S} \circ \mathcal{F}_{f}^{N_{c}}(\mathbf{z^\prime}_r), \\
\text{output gate: }& \boldsymbol{G}_{o}=\boldsymbol{\varphi}_{S} \circ \mathcal{F}_{o}^{N_{c}}(\mathbf{z^\prime}_r), \\
\text{internal state: }& \boldsymbol{s}_{r}=\boldsymbol{G}_{f} \odot \boldsymbol{s}_{r-1}+\boldsymbol{G}_{i} \odot\left(\boldsymbol{\varphi}_{T} \circ \mathcal{F}_{is}^{N_{c}}(\mathbf{z^\prime}_r)\right), \\
\text{output: }& \mathbf{h}_{r+1} = \boldsymbol{G}_{o} \circ \boldsymbol{\varphi}_{T}\left(\boldsymbol{s}_{r}\right),
\end{split}    
\end{align}
where $\mathbf{z^\prime}_r \in \mathbb{R}^{\mathbf{k}}$ is row $r$ of $\mathbf{r}$ rows in matrix $\mathbf{Z}^\prime$. The above set of operations are \emph{unrolled} in the \emph{PCA dimension} to allow for the effect of history. At this point, we have interpreted data in the PCA dimension to be sequential, which aligns with the well-known variance-ordered nature of principal components obtained from snapshot data. We may now use this directional information to predict the dynamics at all future time steps at once (contained in the vectors $\mathbf{h}_0, \mathbf{h}_1, \hdots, \mathbf{h_r}$). This is represented as
\begin{align}
    \mathbf{Z_{NA}} = \varphi_T (\mathcal{F}_{op}^\mathbf{N-k} [ \mathbf{h_{1}}, \mathbf{h_{2}}, \hdots, \mathbf{h_{r+1}} ]),
\end{align}
where the double brackets imply column vector concatenation. The final output of this framework, $\mathbf{Z_{NA}} \in \mathbb{R}^{\mathbf{N-k} \times \mathbf{r}}$, contains all the information of the evolution of the $\mathbf{r}$-dimensional state for all future time steps $\mathbf{N-k}$. Note that the length of $\mathbf{h}_r \in \mathbb{R}^{N_c}$ is a function of the number of hidden-layer neurons $N_c$ in this LSTM cell. In a manner similar to that demonstrated in the autoregressive case, parameter information $w$ is concatenated into the gating (i.e., PCA) dimension. We then extend the implementation of the NA-LSTM-P formulation allowing for a bidirectional gating mechanism in the PCA dimension \cite{graves2005framewise}. Through the use of this augmentation, any output $\mathbf{h}_{r}$ is affected by all inputs $\mathbf{z}^{\prime}_1, \mathbf{z}^{\prime}_2, \hdots, \mathbf{z}^{\prime}_\mathbf{r}$. This decision is physics-informed due to the common knowledge of energy interchange between the different frequencies (and correspondingly, the different PCA basis vectors) in multiscale nonlinear dynamical systems \cite{wang2012proper}. We denote this method by NA-BLSTM-P. }


The \emph{sequential in PCA space} interpretation of the dataset also allows for the use of a comparable architecture given by the one-dimensional temporal convolution network \cite{oord2016wavenet}, where convolutions are performed on the parametric information and the PCA coefficients comprising the $\mathbf{k}$ input timesteps. The reader may find an excellent review of the temporal convolutional network and its applications for reduced-order modeling in \cite{xu2019multi}. As in the previous case, we then obtain a sequence of PCA coefficients for all future timesteps as the output of the network. We denote this method by NA-TCN. As a baseline we also use a fully connected network for predicting from the input timesteps and PCA coefficients. In essence, the PCA coefficients for all input timesteps are flattened and concatenated with parameter information to obtain an input signal that is used to directly predict a flattened vector of PCA coefficients. We denote this method by NA-MLP. To summarize this section, given a burn-in sequence of $\mathbf{k}$ inputs and control variables $w$, these frameworks produce the PCA coefficients for all future timesteps directly. 

\section{Experiments}
\label{Experiments}

We describe the data generation methodology from the inviscid shallow water equations of an advection-dominated system and present the experimental results comparing different learning methods.

\subsection{Data generation for training and inference}

\RM{The inviscid shallow water equations belong to a prototypical system of equations for geophysical flows. In particular, the shallow water equations admit solutions where advection dominates dissipation and poses challenges for conventional ROMs \cite{wang2012proper}. These governing equations are given by
\begin{align}
    \frac{\partial(\rho \eta)}{\partial t}+\frac{\partial(\rho \eta u)}{\partial x}+\frac{\partial(\rho \eta v)}{\partial y} =0  \label{eq1} \\
    \frac{\partial(\rho \eta u)}{\partial t}+\frac{\partial}{\partial x}\left(\rho \eta u^{2}+\frac{1}{2} \rho g \eta^{2}\right)+\frac{\partial(\rho \eta u v)}{\partial y} = 0 \label{eq2} \\
    \frac{\partial(\rho \eta v)}{\partial t}+\frac{\partial(\rho \eta u v)}{\partial x}+\frac{\partial}{\partial y}\left(\rho \eta v^{2}+\frac{1}{2} \rho g \eta^{2}\right) = 0, \label{eq3}
\end{align}
where $\eta$ corresponds to the total fluid column height, and $(u,v)$ is the fluid's horizontal flow velocity, averaged across the vertical column, $g$ is acceleration due to gravity, and $\rho$ is the fluid density, typically set to 1.0. Here, $t,x$ and $y$ are the independent variables: time and the spatial coordinates of the two-dimensional system. Equation \ref{eq1} captures the law of mass conservation, whereas Equations \ref{eq2} and \ref{eq3} denote the conservation of momentum. The initial conditions of the problem are given by 
\begin{align}
    \rho \eta (x,y,t=0) &= e^{-\left(\frac{(x-\bar{x})^2}{2(5e+4)^2} + \frac{(y-\bar{y})^2}{2(5e+4)^2}\right)} \\
    \rho \eta u(x,y,t=0) &= 0 \\
    \rho \eta v(x,y,t=0) &= 0,
\end{align}
}
i.e., a Gaussian perturbation at a particular location on the grid $[\bar{x},\bar{y}] \equiv w$. We solve the system of equations until $t=0.5$ with a time step of 0.001 seconds on a square two-dimensional grid with 64 collocation points to completely capture the advection and gradual decay of this perturbation. Note that these numbers may vary according to the forecasting and fidelity requirements of a particular problem and perturbation. An additional challenge is introduced when we seek to build predictive models solely from observations of $\rho \eta$ conditioned on $w$ mimicking a real-world scenario where complete observations of all relevant variables (in this case, velocities) are unavailable. Equation-based models are thus impossible to construct because of the absence of information from the other variables of the partial differential equation. 

\begin{figure}[h!]
    \centering
    \includegraphics[width=0.8\textwidth]{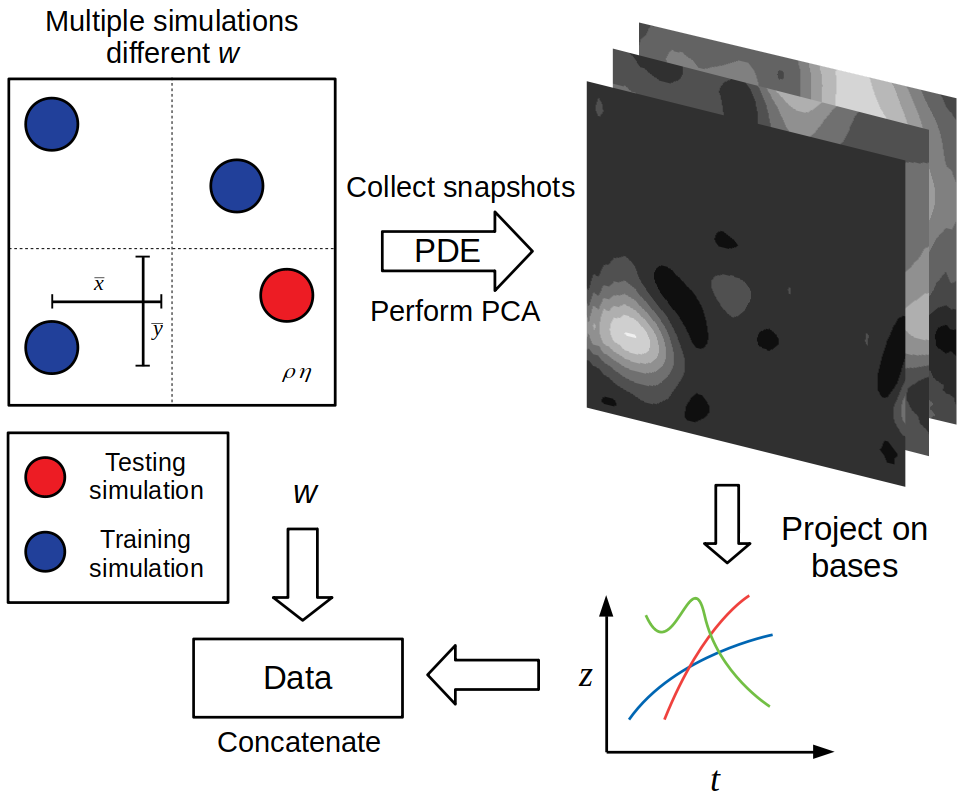}
    \caption{Schematic outlining the generation of training and testing data for the forecasting problem. Multiple simulations are used to generate training and testing snapshots. The training snapshots are used to compute a set of PCA bases on which individual snapshots are projected to obtain time-varying PCA coefficients. These coefficients are predicted for the testing simulation by a data-driven method. Here $z$, $t$, and $w \equiv [\bar{x}, \bar{y}]$ are the PCA coefficients, time, and control parameters, respectively.}
    \label{Fig_schematic}
\end{figure}

Five hundred snapshots of $\rho \text{ and } \eta$ each are generated from 20 different vectors $w$ obtained by Latin hypercube sampling. These snapshots are used to obtain the global PCA bases that span all these simulations. The PCA bases constructed from field (or image) snapshot data capture information in a linear least-squares sense \cite{berkooz1993proper}. Linear combinations of PCA bases may be used to reconstruct the dynamics of the partial differential equation with the use of projection methods. A schematic detailing the generation of training and testing data is shown in Figure \ref{Fig_schematic}. 

Instead of utilizing all possible PCA bases, because of issues of computational complexity (20 simulations $\times$ 500 time steps = 10,000 components), only $\mathbf{r}=40$ PCA bases corresponding to the most energetic structures in the data are retained. Therefore only $\mathbf{r}$ coefficients are required in order to reconstruct a field, given these bases. We note that the choice to set $\mathbf{r}=40$ coefficients is made by assessing the reconstruction error in the transformed bases. This number allows us to reconstruct the original solution with root mean square errors (RMSEs) around the order of 1e-5 and captures more than 95\% of the total variance by measuring the magnitude of singular values in the PCA decomposition. By interpreting that the nonlinear dynamics of our problem may be spanned by the PCA bases in both space and time, we obtain coefficients conditioned on $w$ as well as time. This study seeks to learn the underlying trends of their evolution in time and assesses this learning for different time-series prediction methods. 
For testing, assessments are made on an \emph{unseen} $w$, and a successful reconstruction of the dynamics for these parameters implies that a viable surrogate has been obtained. Moreover, we assume that a short duration of observations, $\mathbf{z_1}, \mathbf{z_2}, \hdots, \mathbf{z_k}$, is available (corresponding to a first window of inputs to our methods). This burn-in window $\mathbf{k}$ corresponds to a very small observation set compared with the full 500 timesteps. We set $\mathbf{k}=20$ for all experiments. This choice was determined through manual experimentation with the objective of having the shortest burn-in duration. We observed that values less than $\mathbf{k}=20$ led to very high deterioration in the results for each method. For training, the time sequence of each PCA component is scaled by the minimum and maximum value of all its counterparts using only the training dataset. We also note that the data generated from these finite-volume simulations can be interpreted to be images on a square-grid. While this lends to the use of convolutional neural networks for modeling dynamics \cite{thuerey2019deep,kim2019deep}, the choice of the PCA projection allows for interpretability of the evolving coefficients in terms of physically relevant spectral content.

\subsection{Experimental setup}

All the learning methods investigated here utilized only two hidden layers to allow for low training and inference times. All our data generation and trained model assessments used Python 3.6.8 and TensorFlow 1.14.0 on an Ubuntu 18.04 operating system with 8 GB of RAM. All the source code for this study, including training and testing data, is available at \texttt{https://github.com/rmjcs2020/NATSurrogates}.

\begin{table*}[ht]
\small
\centering
\caption{High-performing hyperparameter values obtained by DeepHyper and corresponding training, test, and final-time reconstruction errors. The proposed method NA-BLSTM-P can be seen to provide the lowest mean squared error (MSE) for this problem.}
\begin{tabular}{|c|c|c|c|c|c|c|}
\hline
Method     & Neurons & Batch size & Learning rate & \begin{tabular}[c]{@{}c@{}}Training loss\\ MSE\end{tabular} & \begin{tabular}[c]{@{}c@{}}Testing loss\\ MSE/Variance\end{tabular} & \begin{tabular}[c]{@{}c@{}}Field error\\ MSE\end{tabular} \\ \hline
GP         & N/A     & N/A        & N/A           & N/A                                                         & 2.85e-3 / 6.15e-5                                                   & 0.000146                                                  \\ \hline
A-LSTM-T   & 88      & 152        & 0.000258      & 0.000503                                                    & 2.26e-1 / 4.59e-3                                                   & 0.000794                                                  \\ \hline
A-LSTM-T-R & 29      & 178        & 0.651         & 0.0187                                                      & 8.77e-3 / 5.61e-4                                                   & 0.000352                                                  \\ \hline
NA-LSTM-T  & 102     & 6          & 0.0342        & 0.00398                                                     & 1.79e-3 / 2.15e-5                                                   & 5.77E-05                                                  \\ \hline
NA-LSTM-P  & 88      & 9          & 0.000258      & 0.00235                                                     & 1.41e-3 / 1.26e-5                                                   & 4.12E-05                                                  \\ \hline
NA-BLSTM-P & 145     & 6          & 0.00523       & 0.00281                                                     & \textbf{1.26e-5 / 5.23e-6}                                                   & \textbf{3.96E-05}                                                  \\ \hline
NA-TCN     & 145     & 6          & 0.00523       & 0.00455                                                     & 2.10e-3 / 2.02e-5                                                   & 4.33E-05                                                  \\ \hline
NA-MLP     & 72      & 4          & 0.00184       & 0.00477                                                     & 1.36e-3 / 1.18e-5                                                   & 5.23E-05                                                  \\ \hline
\end{tabular}
\label{Table2}
\end{table*}

An asynchronous hyperparameter optimization of the methods that we consider was performed on 256 Intel Knights Landing compute nodes for 6 hours with the best models being chosen by their \RM{validation} loss. Subsequent evaluations of the trained models were performed on an eighth-generation Intel Core-I7 processor. For optimizing the hyperparameters of the different methods, we use DeepHyper \cite{balaprakash2018deephyper}, a hyperparameter search package that leverages Bayesian black-box optimization from scikit-optimize \cite{pedregosa2011scikit}. This also allows for more confident conclusions about the results of our models and removes bias due to default or manual parameter settings. 

The search space for all hyperparameters was limited to the number of neurons (for each of two hidden layers), the batch size, and the learning rate. A smaller search space ensured more model evaluations to allow for effective search-space exploration. Our search-space bounds are [10,200] for the hidden-layer neurons and [1e-4,1e-1] for the learning rates. Batch size bounds were set according to the choice of method: autoregressive methods have batch sizes between 64 and 256, and non-autoregressive methods have batch sizes between 1 and 10.  Two hyperparameter optimization experiments are performed for each method in this study. The first utilizes a standard training, and the second incorporates dropout regularization \cite{srivastava2014dropout} for stability analyses of methods. Both searches utilize the same bounds. On average, each method was evaluated for more than 400 unique hyperparameter configurations. The high-performing hyperparameters obtained by using DeepHyper are shown in Table \ref{Table2}. Models were trained with convergence callbacks based on training loss (less than 1e-5 RMSE) or training time out (corresponding to 5,000 epochs). 

\subsection{Comparison of different methods}
\label{deterministic}

Here, we compare the different autoregressive and non-autoregressive learning methods  with GP and show that non-autoregressive methods are superior to other methods with respect to forecasting accuracy and stability.

The different methods are compared based on the MSE of PCA coefficient predictions for the test simulation. We also show error variance for each method. The training loss, testing error, and field reconstruction error results are shown in Table \ref{Table2}. First, we compare the autoregressive variants. We observe that both A-LSTM-T and the A-LSTM-T-R perform poorly. \RM{While the latter exhibits a saturation after a certain prediction horizon, the former displays compounding errors that diverge from the true trajectory. These results are reflected in their test MSEs of 2.26e-1 and 8.77e-3, the two highest in our set of experiments. We also note that A-LSTM-T shows a very low training loss of around 5.03e-4, which is far lower than those obtained by the other methods; however, its testing performance is poor, hinting at issues of stability for long-term feedback predictions.}

Next we look at the non-autoregressive methods. \RM{For these methods, testing MSEs show that the proposed method NA-BLSTM-P has the lowest magnitude of error (1.26e-5) compared with NA-LSTM-T, NA-LSTM-P,  NA-TCN, and NA-MLP (1.79e-3, 1.41e-3, 2.10e-3, and 1.36e-3, respectively). This may be attributed to the physics-aware nature of the gating in PCA space where information from multiple PCA coefficients interacts globally across the $\mathbf{r}=40$ components. The gating in PCA space has a positive effect on the performance of the non-autoregressive methods, with both NA-LSTM-P and NA-BLSTM-P showing the two best test results. The sequential nature of the ordered PCA coefficients is thus leveraged successfully. In terms of training losses, NA-LSTM-P has the lowest training loss among non-autoregressive methods of 2.35e-3, whereas  NA-BLSTM-P has the next best (but comparable) training loss of 2.81e-3. These are followed by NA-LSTM-T (3.98e-3), NA-TCN (4.55e-3), NA-MLP (4.77e-3). As expected, non-autoregressive methods exhibit lower testing errors than their autoregressive counterparts.}

\begin{figure*}[ht]
    \centering
    \includegraphics[width=0.95\textwidth]{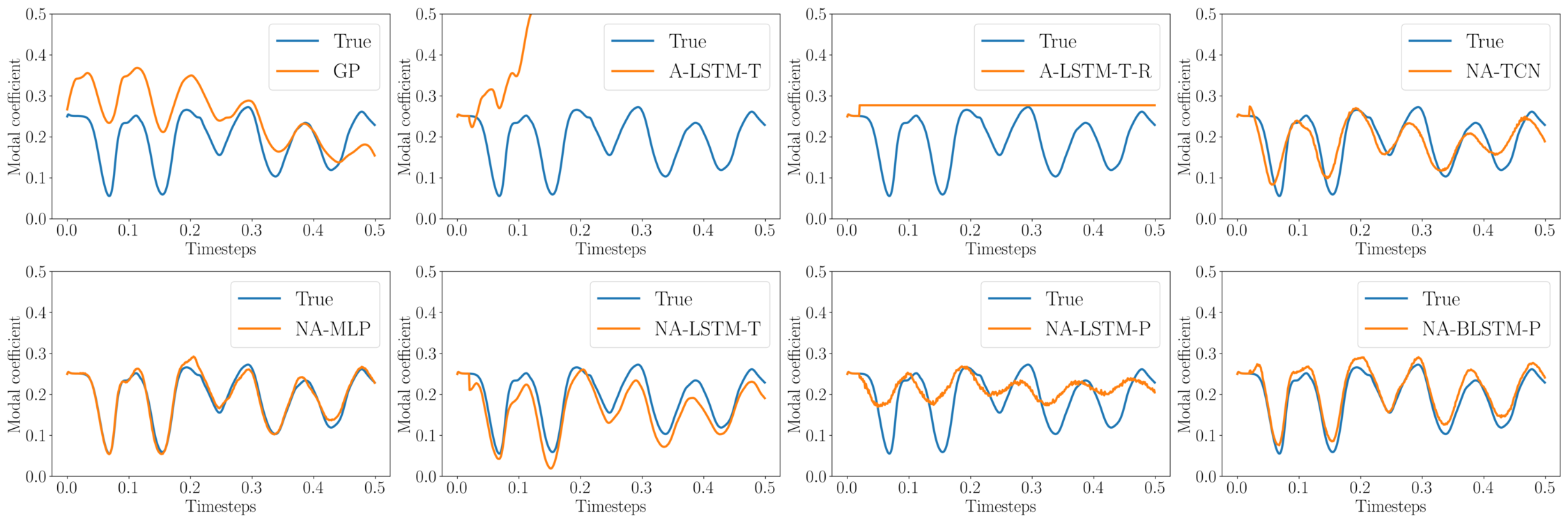}
    \caption{Predictive ability for the assessed frameworks for PCA component 1. Non-autoregressive methods are seen to be better than their autoregressive counterparts. Note that in order to build a model, GP requires the solution of a partial differential equation in addition to greater observations from the true system.}
    \label{Fig_1a}
\end{figure*}

\begin{figure}[h!]
    \centering
    \includegraphics[width=\textwidth]{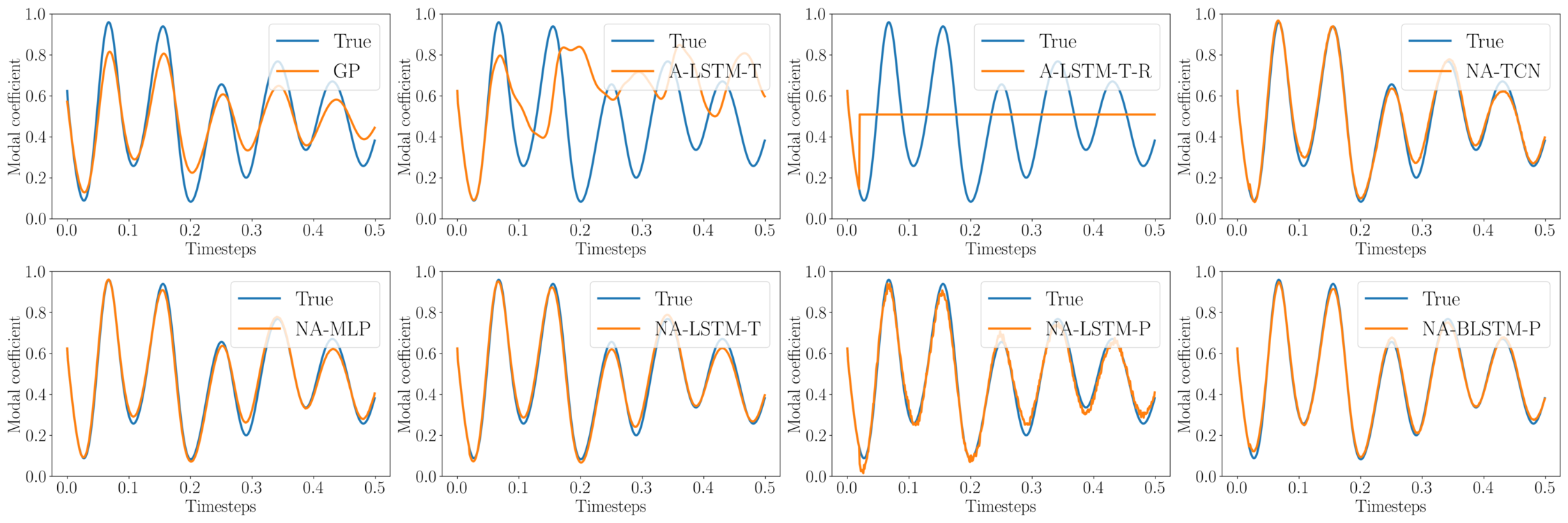}
    \caption{Predictive ability for the assessed frameworks for PCA component 2. Non-autoregressive methods are seen to be better than their autoregressive counterparts. Note that in order to build a model, GP requires the solution of a partial differential equation in addition to greater observations from the true system.}
    \label{Fig_2a}
\end{figure}

\begin{figure}[h!]
    \centering
    \includegraphics[width=\textwidth]{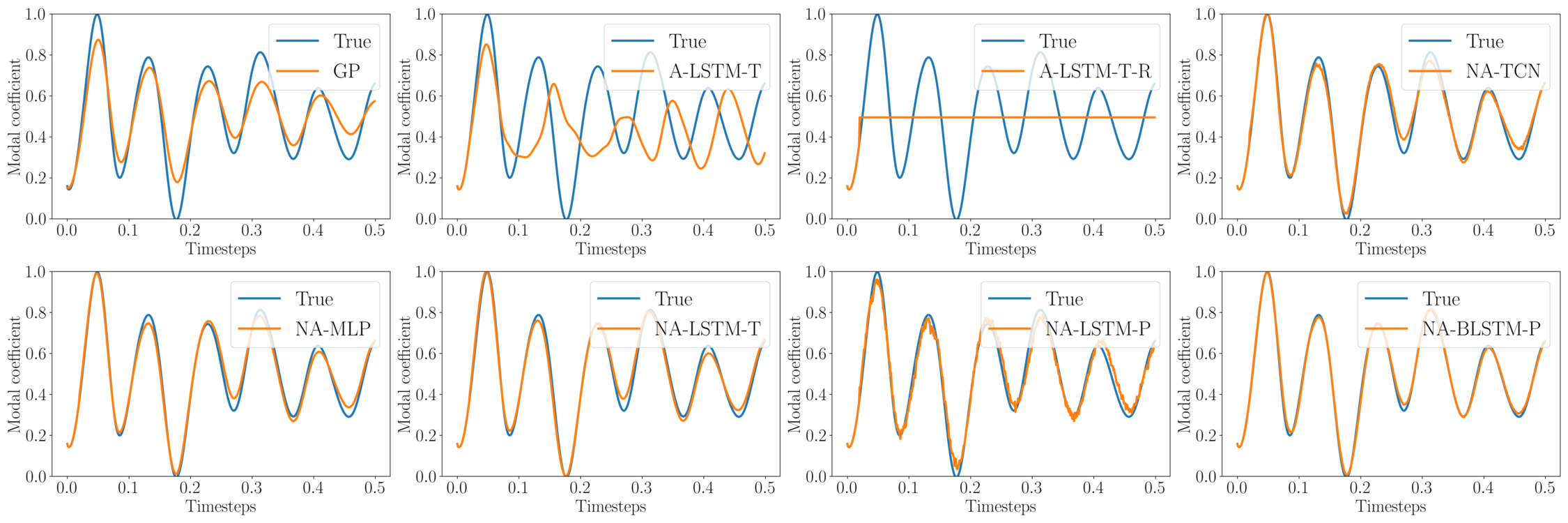}
    \caption{Predictive ability for the assessed frameworks for PCA component 3. Non-autoregressive methods are seen to be better than their autoregressive counterparts. Note that in order to build a model, GP requires the solution of a partial differential equation in addition to greater observations from the true system.}
    \label{Fig_3a}
\end{figure}

\begin{figure}[h!]
    \centering
    \includegraphics[width=\textwidth]{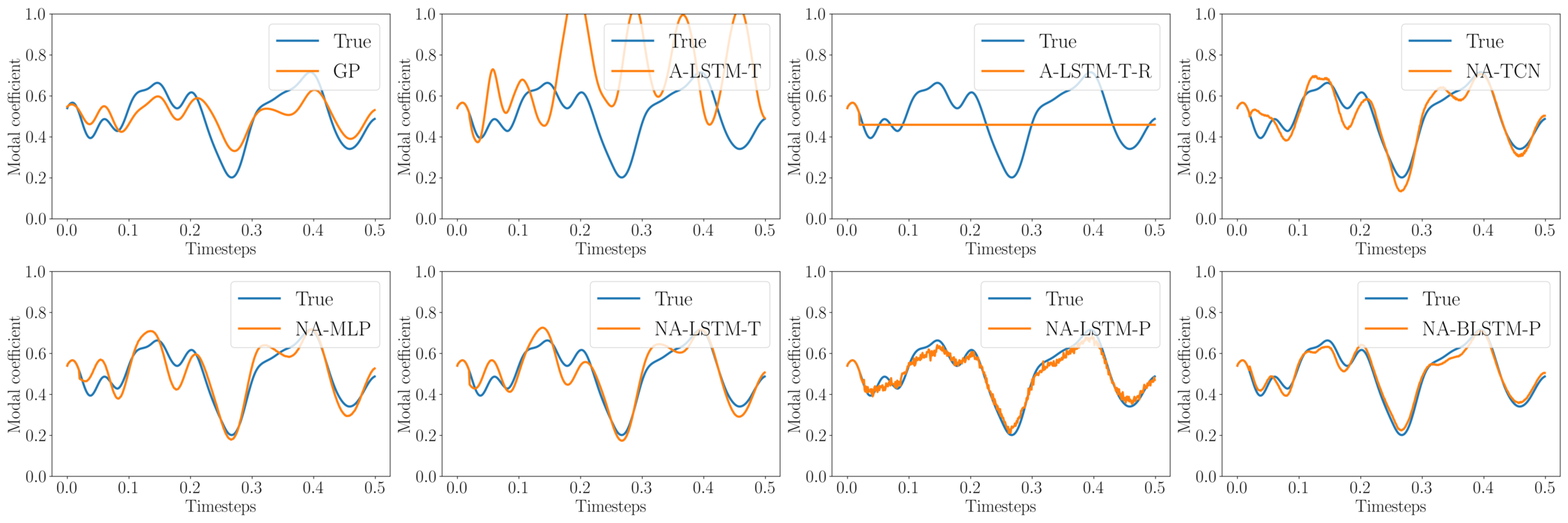}
    \caption{Predictive ability for the assessed frameworks for PCA component 4. Non-autoregressive methods are seen to be better than their autoregressive counterparts. Note that in order to build a model, GP requires the solution of a partial differential equation in addition to greater observations from the true system.}
    \label{Fig_4a}
\end{figure}

We also note that all the non-autoregressive methods display lower testing errors than GP does (which has a testing error of 2.85e-3). The  comparison is made more favorable by the fact that GP requires an equation-based evolution of all variables in the system whereas our data-driven methods are built solely for $\rho \eta$. In addition, GP  requires the PCA of the other variables to identify the latent space for evolution. These results are displayed in Figure \ref{Fig_1a}, which shows testing predictions for the normalized coefficients of the first PCA coefficient for $\rho \eta$. Similar results are obtained for higher-order coefficients as seen in Figures \ref{Fig_2a}, \ref{Fig_3a} and \ref{Fig_4a}.

\begin{figure}[h!]
    \centering
    \includegraphics[width=\textwidth]{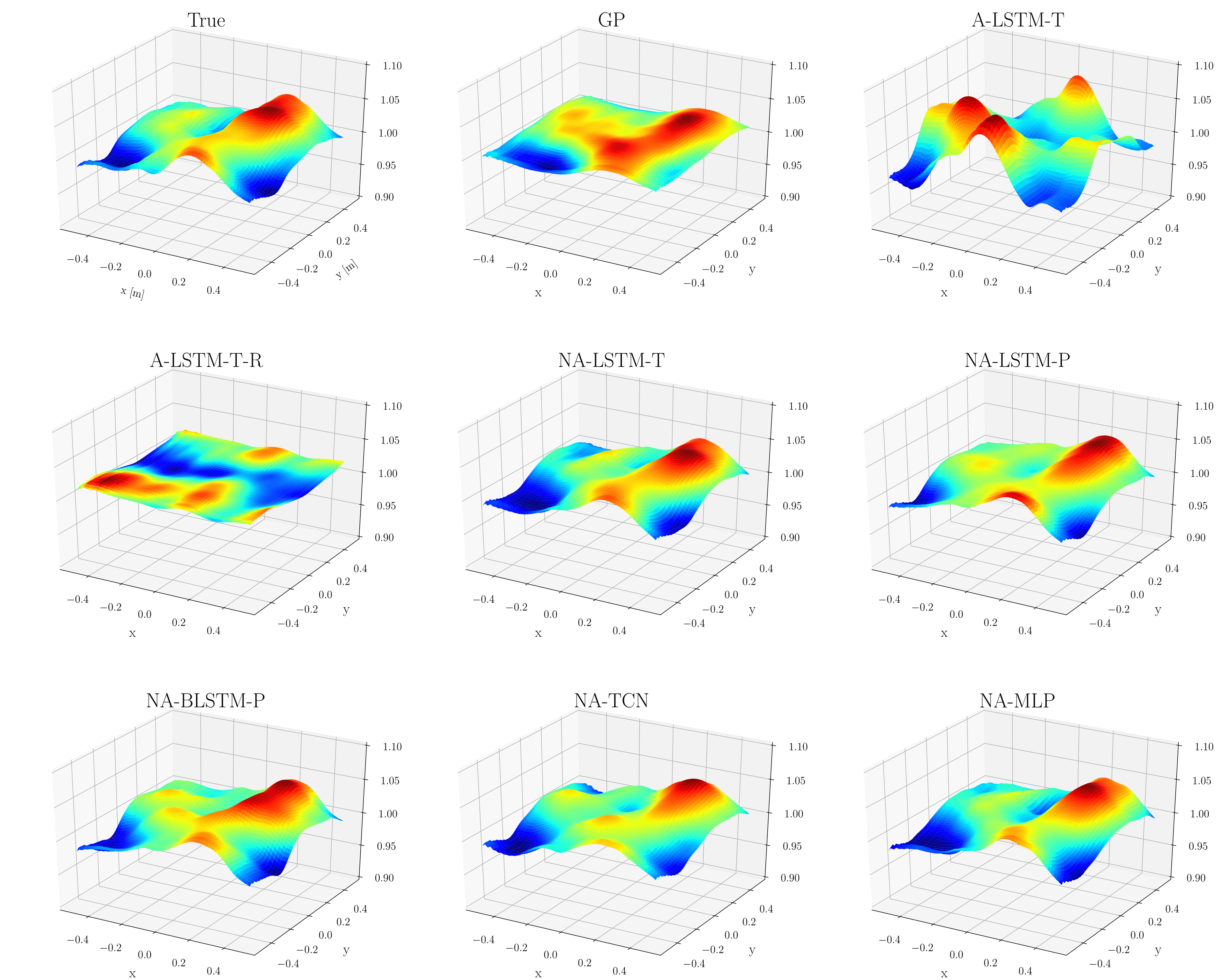}
    \caption{Predicted fields at final time. We remind the reader that GP requires the solution of a partial differential equation in addition to greater observations from the true system to build a model.}
    \label{Fields}
\end{figure}

We plot final time fields of $\rho \eta$ in Figure \ref{Fields}. Field reconstructions, compared to the truth, show that final time predictions of non-autoregressive frameworks are more successful in stable predictions. Observe, that GP proves less accurate than the NAT methods in addition to requiring an equation-based evolution of \emph{all} variables via a set of coupled ODEs.

\subsection{Stability analysis}

\begin{table*}[ht]
\small
\centering
\caption{Optimal hyperparameters obtained by DeepHyper with dropout during training and inference. Non-autoregressive methods are seen to perform better than their autoregressive counterparts. In addition, the proposed method (NA-BLSTM-P) can be seen to provide the lowest mean squared error (MSE) for this problem.}
\begin{tabular}{|c|c|c|c|c|c|c|}
\hline
Method     & Neurons & Batch size & Learning rate & \begin{tabular}[c]{@{}c@{}}Training loss\\ MSE\end{tabular} & \begin{tabular}[c]{@{}c@{}}Testing loss\\ MSE/Variance\end{tabular} & \begin{tabular}[c]{@{}c@{}}Trainable\\ parameters\end{tabular} \\ \hline
GP         & N/A     & N/A        & N/A           & N/A                                                         & 2.85e-3 / 6.15e-5                                                   & N/A                                                            \\ \hline
A-LSTM-T   & 197     & 186        & 2.68E-02      & 2.11E-01                                                    & 1.43e-2 / 1.42e-3                                                   & 508300                                                         \\ \hline
A-LSTM-T-R & 101     & 217        & 7.67E-01      & 1.86E-02                                                    & 1.39e-1 / 9.24e-3                                                   & 144472                                                         \\ \hline
NA-LSTM-T  & 145     & 6          & 5.23E-03      & 4.23E-03                                                    & 1.32e-3 / 1.06e-5                                                   & 3081020                                                        \\ \hline
NA-LSTM-P  & 88      & 9          & 2.58E-04      & 3.23E-03                                                    & 1.96e-3 / 2.06e-5                                                   & 143392                                                         \\ \hline
NA-BLSTM-P & 145     & 6          & 5.23E-03      & 2.37E-03                                                    & \textbf{8.38e-4 / 5.71e-6}                                                   & 838000                                                         \\ \hline
NA-TCN     & 88      & 9          & 2.58E-04      & 4.94E-03                                                    & 1.86e-3 / 1.26e-5                                                   & 504544                                                         \\ \hline
NA-MLP     & 119     & 4          & 9.62E-04      & 4.73E-03                                                    & 1.53e-3 / 1.44e-5                                                   & 2413837                                                        \\ \hline
\end{tabular}
\label{Table3}
\end{table*}

\RM{Analyzing the stability of ROM models is a critical step in the development and deployment of such models. To that end, we perturb the model through the incorporation of dropout \cite{srivastava2014dropout} and analyze its stability with respect to these perturbations. Dropout is a well-known machine learning regularization technique where randomly selected neurons in a neural network are set to zero during training. This prevents a neural network from overfitting. Recently, dropout has also been used to represent model uncertainty in deep learning \cite{gal2016dropout}, where random neurons are switched off during several predictions. Moments may then be generated from these multiple predictions. In this section, we shall use dropout during training to ascertain the effect of regularization on our forecast methods. Dropout during inference will be utilized to analyze the sensitivity of the surrogate models to error accumulation. Our hypothesis is that greater sensitivity to perturbations of the machine-learned models will lead to greater forecasting error. This analysis is also motivated by the fact that few reduced-order modeling techniques include notions of model-form uncertainty or sensitivity analyses during forecasting.}

We perform the stability analysis with the incorporation of dropout during both training and inference with the former aimed at avoiding overfitting. The purpose of incorporating dropout during inference is to assess the effect of perturbations on the trained frameworks. This assessment can provide further validation of the conclusions made in Section \ref{deterministic}, where low training losses for autoregressive methods were not correlated with good testing performance. We believe that the error growth over a long-term prediction horizon was the cause of model inaccuracy in testing. Predictions obtained by using dropout at inference time allow for multiple time-series predictions by a trained network. A slight perturbation to the model (by switching off a subcomponent) would lead to error accumulation for autoregressive methods whereas the non-autoregressive methods would bypass this effectively. We therefore use this technique to perform a stability analysis of the trained networks.

\begin{figure}[ht]
    \centering
    \includegraphics[width=0.95\textwidth]{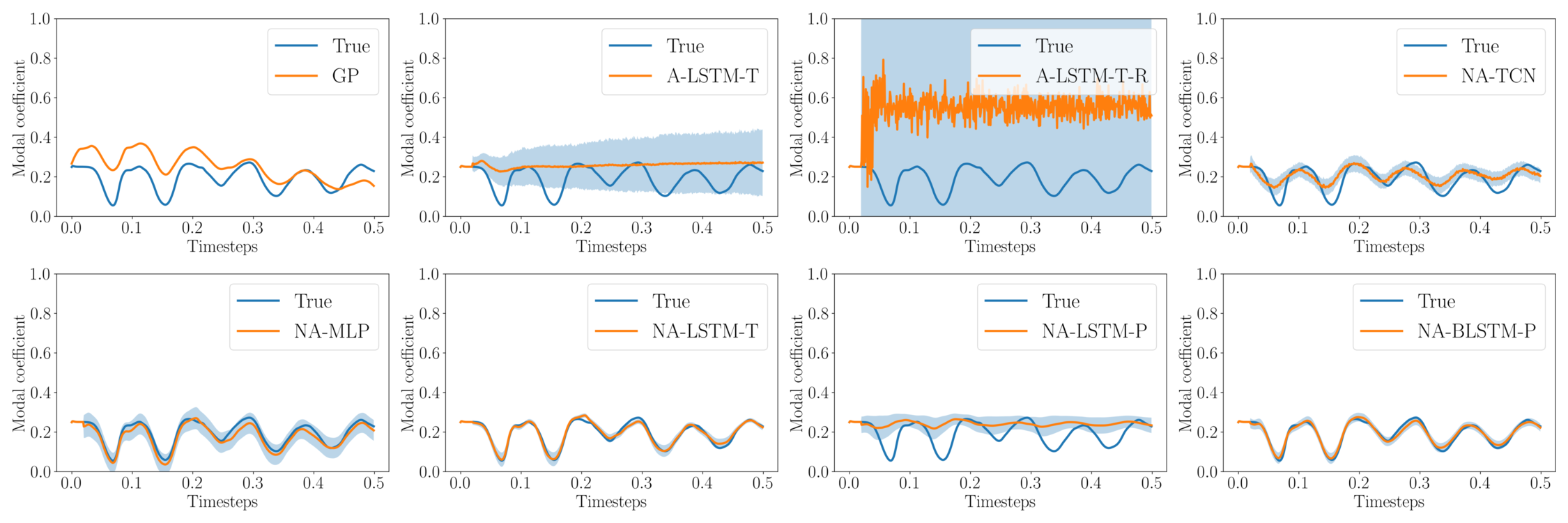}
    \caption{Predictive ability for the assessed frameworks for PCA component 1. Note that GP requires the solution of a partial differential equation in addition to complete observations from the true system to build a model. In addition, GP is deterministic.}
    \label{Fig_1_do}
\end{figure}

\begin{figure}[ht]
    \centering
    \includegraphics[width=0.95\textwidth]{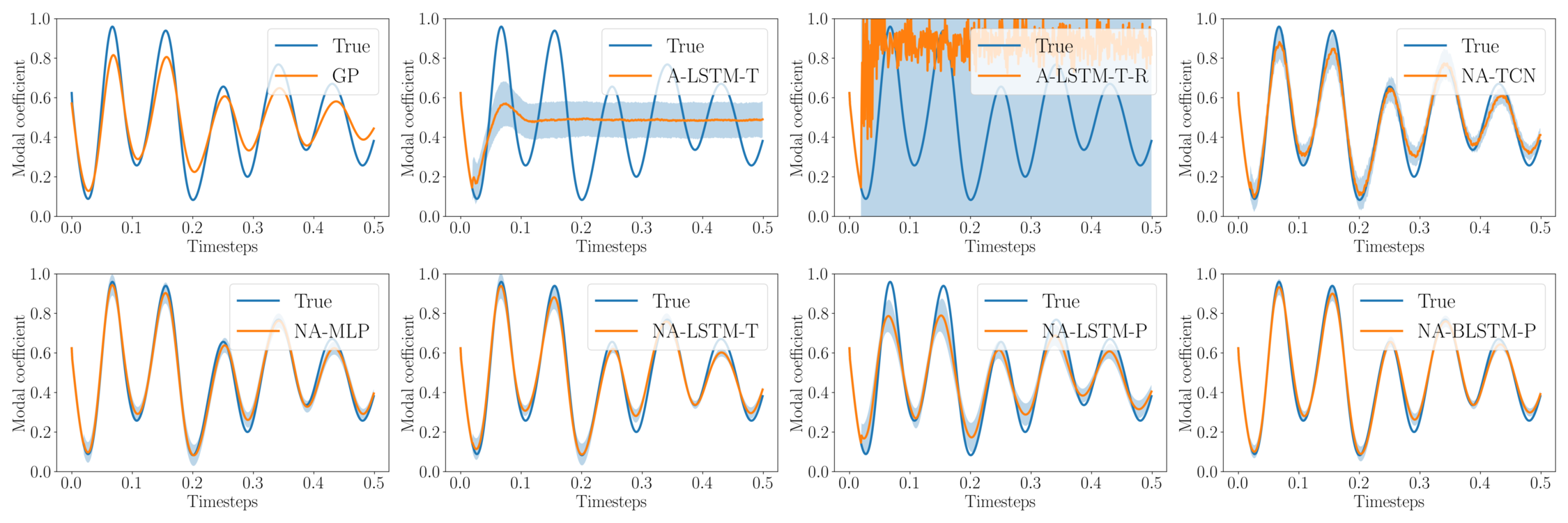}
    \caption{Predictive ability for the assessed frameworks for PCA component 2. Note that GP requires the solution of a partial differential equation in addition to complete observations from the true system to build a model. In addition, GP is deterministic.}
    \label{Fig_2_do}
\end{figure}

\begin{figure}[ht]
    \centering
    \includegraphics[width=0.95\textwidth]{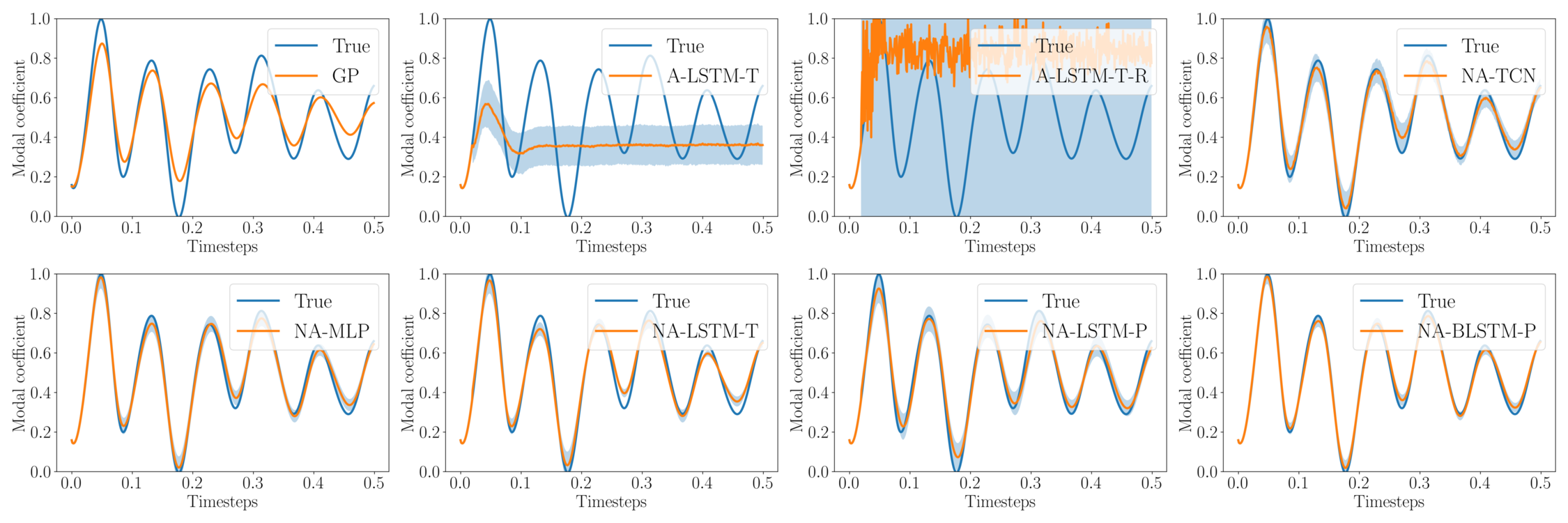}
    \caption{Predictive ability for the assessed frameworks for PCA component 3. Note that GP requires the solution of a partial differential equation in addition to complete observations from the true system to build a model. In addition, GP is deterministic.}
    \label{Fig_3_do}
\end{figure}

\begin{figure}[ht]
    \centering
    \includegraphics[width=0.95\textwidth]{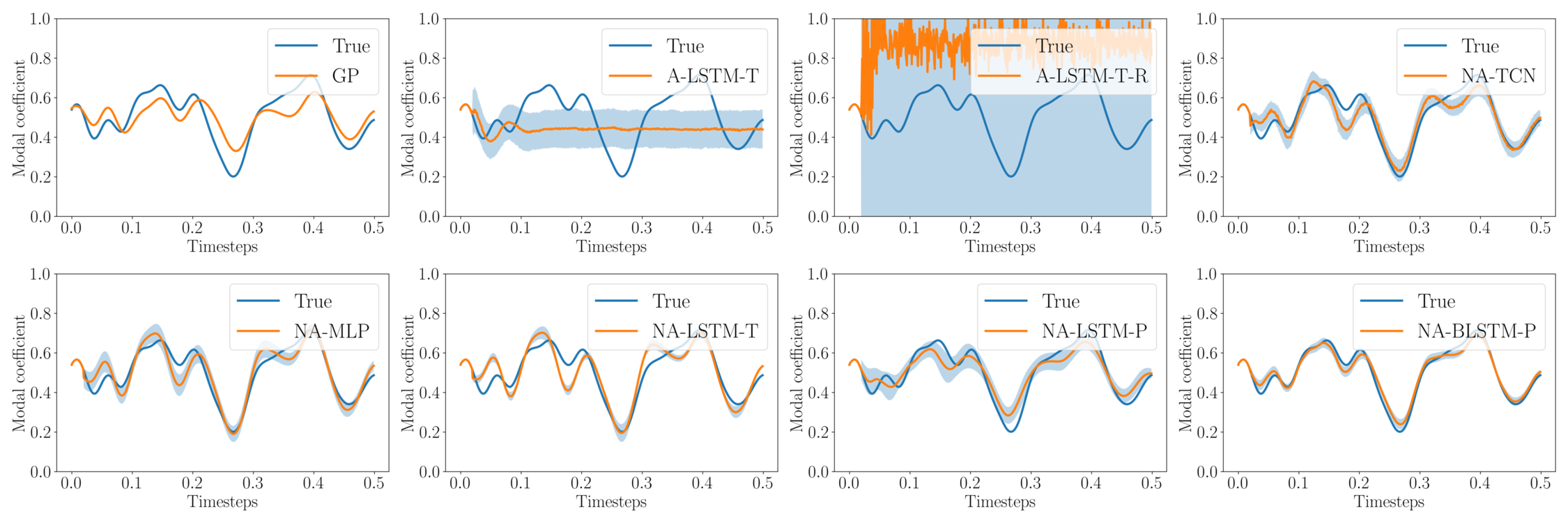}
    \caption{Predictive ability for the assessed frameworks for PCA component 4. Note that GP requires the solution of a partial differential equation in addition to complete observations from the true system to build a model. In addition, GP is deterministic.}
    \label{Fig_4_do}
\end{figure}

We run a hyperparameter search for all the learning methods with dropout for each hidden layer. As a default, we used a dropout probability of $0.2$. Results are shown in Table \ref{Table3}. Once trained, each method is tasked with predicting on the test dataset 1,000 times. Then the mean of all these predictions and their variance are calculated. The mean of the predictions is then compared with the true data (which is deterministic) to obtain testing errors. Figure \ref{Fig_1_do} shows the predicted mean and variance of these 1,000 inferences for all assessed methods on the first principal component.
Results for this analysis on other higher-order PCA components are shown in Figures \ref{Fig_2_do}, \ref{Fig_3_do}, \ref{Fig_4_do}. Note that the GP method is equation-based and \emph{deterministic}. \RM{The A-LSTM-T method saturates as predictions evolve in time. However, that A-LSTM-T-R cannot match the right phase or frequency of the oscillations for any PCA coefficients and generates a random noisy signal. In contrast, the NA methods are able to match the phase of the predictions appropriately. These trends are repeated for the other coefficients as well. Quantitative comparisons across all coefficients are given in Table \ref{Table3} and show that the proposed methods NA-BLSTM-P and NA-LSTM-P outperform their counterparts. We also show the number of trainable parameters for each architecture in this table. One can note that the proposed methods NA-BLSTM-P and NA-LSTM-P require a comparable number of trainable parameters compared to NA-TCN and far fewer than the NA-LSTM-T and NA-MLP methods. We note here, that the latter two algorithms do not interpret the data to be sequential in PCA space.}

\RM{Non-autoregressive methods, in general, provide several orders of magnitude acceleration over GP and the autoregressive methods, thereby proving suitable for their ultimate application in cost reduction. For instance, the average time-to-solution for a non-autoregressive method was around 0.01 seconds, whereas the autoregressive methods required approximately 3 seconds for forecasting. The equation-based GP model required around 20 seconds for a complete simulation. The readers may note that the large cost of the GP model stems from the need to retain 40 principal components which causes greater computational demands due to the complexity of the nonlinear term computation.}

\begin{figure}[ht]
    \centering
    \includegraphics[width=0.95\textwidth]{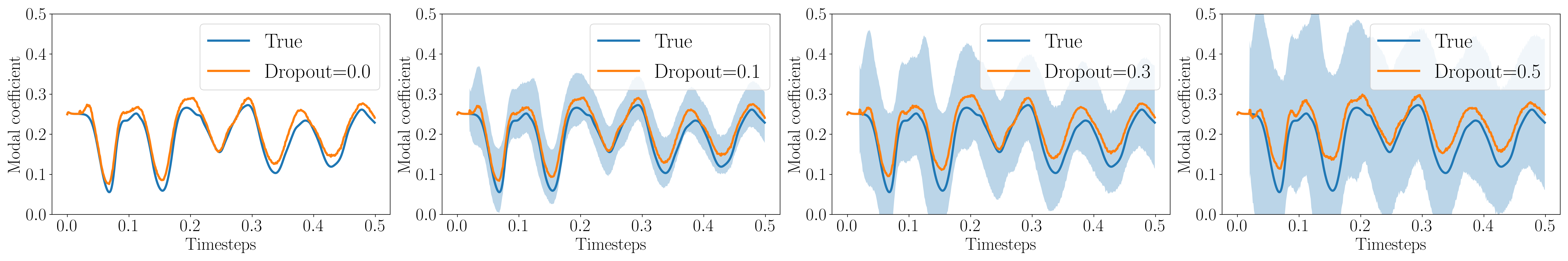}
    \vspace{-2.0mm}
    \caption{Ablation study focusing on the effect of dropout probability on the NA-BLSTM-P method. Testing mean-squared errors / error variance for the four models were (from left to right): 8.40e-5 / 5.23e-6, 8.53e-4 / 5.08e-6, 1.01e-3 / 5.44e-6, and 1.50e-3 / 8.59e-6.}
    \label{Dropout_study}
\end{figure}

For another validation, we take the best hyperparameters obtained for NA-BLSTM-P and perform an ablation study for the magnitude of the dropout probability, as shown in Figure \ref{Dropout_study}. The network was deployed four times with dropout probability values of 0.0, 0.1, 0.3, and 0.5; 1,000 inferences were made on the test dataset for each value of the dropout probability. \RM{As the dropout probabilities of neurons increase during inference, the network is seen to display increased variability. However, mean predictions for the ensemble show good agreement with the underlying truth. We note that this model was trained without dropping out neurons and the role of dropout during inference is to simulate error accumulation.} The viable performance for different dropout magnitudes indicates that the model is robust to error forcing at different magnitudes for the entire length of the forecast, thereby displaying robustness for long-term forecasts.

\subsection{Parametric meta-model for latent space initialization}

\RM{In this section, we assess further possibilities for accelerating our reduced-order model by removing the dependence to the previously introduced burn-in duration. We remind the reader that our experiments, thus far, assume the availability of a slice of the trajectory for a test initial condition (from a numerical solver) to initialize the deployment of the data-driven forecast methods. This represents a potential bottleneck for physical systems where large lead-times may be necessary for the effect of control parameters to cause learnable differences in trajectories. To that end, we propose the use of a meta-modeling strategy that predicts the initial burn-in trajectory from $w$ alone. Essentially, our meta-model learns to predict the trajectory for the first $\textbf{k}$ time steps by observing $w$, following this our time-series method predictions the trajectory for the $\mathbf{N-k}$ time steps.}

\RM{For simplicity, we select a 3-layer perceptron with 20, 40 and 20 neurons in hidden layers 1, 2 and 3 respectively. Each hidden layer is also equipped with a rectified linear activation function. This architecture is used to map from $w$ to $\mathbf{z_1,z_2,\hdots,z_k}$. Our dataset remains unchanged, with 20 training points being used to train this map. Note that the paucity of training data also motivates the use of such a simple framework. We train our framework with an $L_1$-regularization strategy where the objective function of our training is penalized by the sum of absolute values of trainable parameters.}

\begin{figure}
    \centering
    \includegraphics[width=0.8\textwidth]{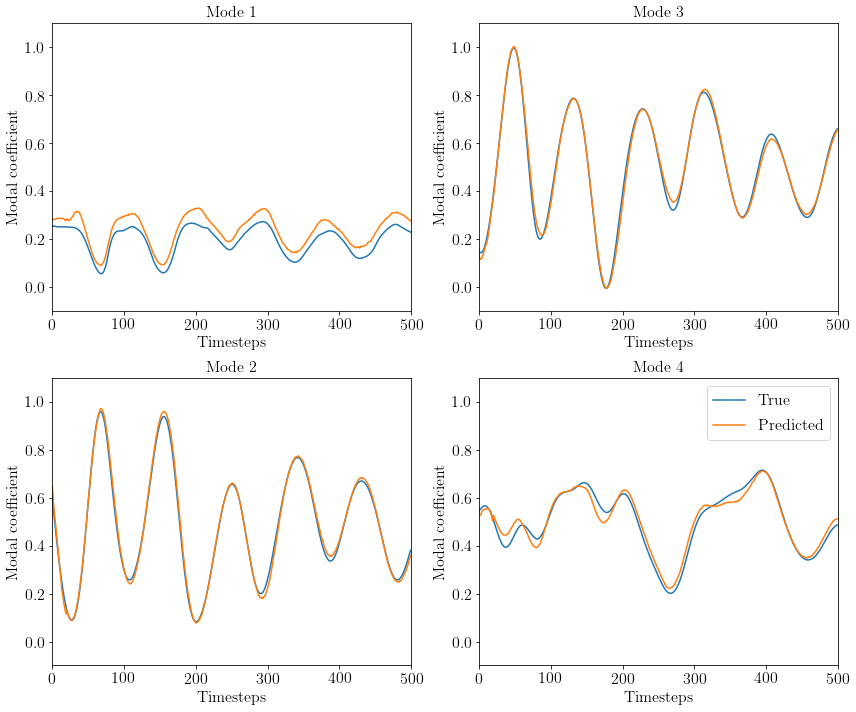}
    \caption{Predictive ability for the NA-BLSTM-P method equipped with a MLP-based meta-model to predict the burn-in sequence given $w$.}
    \label{fig:MM_1}
\end{figure}

\begin{figure}
    \centering
    \includegraphics[width=0.8\textwidth]{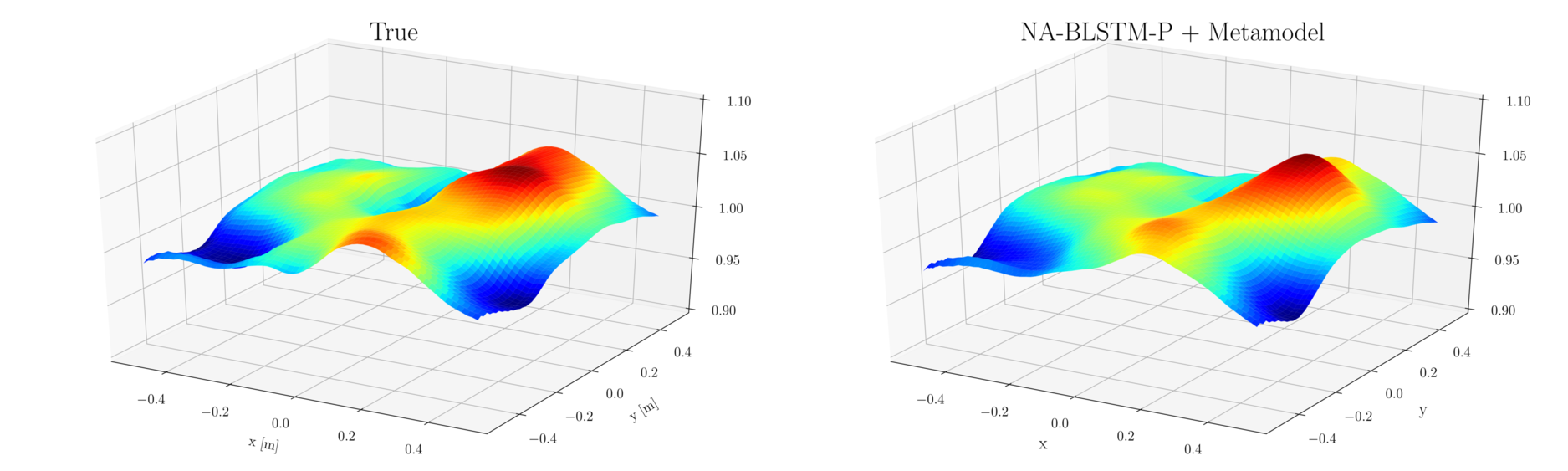}
    \caption{Predictive ability for the NA-BLSTM-P method equipped with a MLP-based meta-model. The meta-model predicts the burn-in sequence given $w$ which may then be used as an input to the non-autoregressive surrogate.}
    \label{fig:MM_2}
\end{figure}

\RM{PCA coefficient predictions from a deployment of the NA-BLSTM-P model trained in Section \ref{deterministic} are shown in Figure \ref{fig:MM_1}. In contrast to the results obtained previously, the initial burn-in window is not available from a numerical simulation but is predicted by our meta-model. One can observe that the combination of the two data-driven maps allows for the direct prediction of a trajectory given solely initial conditions from $w$. Contours for the associated trajectory are shown in Figure \ref{fig:MM_2} where the final time flow-field is reconstructed accurately.}

\section{Conclusions and future work}

We introduce a novel methodology for the accurate prediction of nonlinear dynamical systems solely from data. The dataset is interpreted to be sequential in PCA space due to the ordered nature of the PCA bases. Consequently, this allows for gating in the PCA coefficient dimension. Through this interpretation of the nonlinear dynamics, the introduced methods address challenges related to the stability of traditional time-series learning methods such as the LSTM for advection-dominated problems. 

Our results show that the incorporation of a bidirectional gating in the PCA coefficient dimension leads to the lowest testing and reconstruction errors for the nonlinear dynamics of the shallow water equations. In addition, the choice of the bidirectional gating is \emph{physics-informed} since the PCA coefficients are ordered according to decreasing spectral content capture. In physics, this variance corresponds to the amount of spectral content captured by the different bases. The bidirectional LSTM allows for the output to be influenced by relevant spectral content from multiple PCA bases and may be the cause of the best performance among the methods studied here. We compare these results with other non-autoregressive methods such as the non-autoregressive temporal convolutional network and the non-autoregressive multilayered perceptron where superior performance is observed. The novel interpretation of nonlinear dynamics also allows for rapid inference with orders of magnitude reduction in inference times in comparison with the benchmark GP and autoregressive methods. We also note that the model size of the proposed non-autoregressive bidirectional LSTM is comparatively lower than that of the other non-autoregressive methods. These features are useful for lightweight deployments of the proposed methods for applications in control, data assimilation, and parametric forecasting.

A few limitations of the non-autoregressive methods need to be explored. While these methods aid in bypassing issues of stability and slow inference times, their efficacy for forecasting  (for unseen $t$) remains to be seen. Our current problem is purely transient in nature, and precise prediction beyond 500 timesteps is not desired \RM{to avoid extrapolation}. However, these methods must be assessed for datasets where self-similar temporal information can be leveraged to make forecasts. Another limitation may arise from the nature of training these non-autoregressive methods. Since each nonlinear dynamical system solve is considered a sample, much larger datasets may need to be generated for effective learning. Training times (seen in Table \ref{Table3}) show that these methods may be costly to train on large data sets. Our future work is aimed at addressing these issues.

\section*{Acknowledgments}
This material is based upon work supported by the U.S. Department of Energy (DOE), Office of Science, Office of Advanced Scientific Computing Research, under Contract DE-AC02-06CH11357. This research was funded in part and used resources of the Argonne Leadership Computing Facility, which is a DOE Office of Science User Facility supported under Contract DE-AC02-06CH11357. RM acknowledges support from the Margaret Butler Fellowship at the Argonne Leadership Computing Facility. This paper describes objective technical results and analysis. Any subjective views or opinions that might be expressed in the paper do not necessarily represent the views of the U.S. DOE or the United States Government. Declaration of Interests - None.

\bibliographystyle{unsrt}  
\bibliography{references}  

\end{document}